# Phonons in a one-dimensional microfluidic crystal

**TSEVI BEATUS[1], TSVI TLUSTY[2] AND ROY BAR-ZIV[1]\***

[1]Departments of Materials and Interfaces, Weizmann Institute of Science, Israel
[2]Physics of Complex Systems, Weizmann Institute of Science, Israel
\*e-mail: barzivr@wisemail.weizmann.ac.il

The development of a general theoretical framework for describing the behaviour of a crystal driven far from equilibrium has proved difficult[1]. Microfluidic crystals, formed by the introduction of droplets of immiscible fluid into a liquid-filled channel, provide a convenient means to explore and develop models to describe non-equilibrium dynamics[2–11]. Owing to the fact that these systems operate at low Reynolds number (Re), in which viscous rather than inertial effects dominate the dissipation of energy, vibrations are expected to be over-damped and play little role in their dynamics[12–14]. Against such expectation, we report the emergence of collective normal vibrational modes (equivalent to acoustic 'phonons') in a one-dimensional microfluidic crystal of water-in-oil droplets at Re $\sim 10^{-4}$. These phonons propagate at an ultra-low sound velocity of $\sim$100 µm s$^{-1}$ and frequencies of a few hertz, exhibit unusual dispersion relations markedly different to those of harmonic crystals, and give rise to a variety of crystal instabilities that could have implications for the behaviour of commercial microfluidic systems. First-principles theory shows that these phonons are an outcome of the symmetry-breaking flow field that induces long-range inter-droplet interactions, similar in nature to those observed in many other systems including dusty plasma crystals[15,16], vortices in superconductors[17,18], active membranes[19] and nucleoprotein filaments[20].

To investigate many-body effects of one-dimensional (1D) hydrodynamic crystals, we built a microfluidic water-in-oil droplet generator[2] (Fig. 1a, Methods section and Supplementary Information, Movie S1). Water droplets formed at a T-junction between water and oil channels under continuous flow, emanating at a constant rate with uniform radii $R$ (10–15 µm) and fixed inter-droplet distances $a$ (10–200 µm). The thin channel ($h = 10$ µm) deformed the droplets into discs, confining their motion to 2D and exerting friction with the floor and ceiling (Fig. 1b). Owing to friction, the droplets were dragged by the oil at a velocity $u_{\rm d}$ (150−800 µm s$^{-1}$) that was slower than the oil ($u_{\rm oil} \sim 5u_{\rm d}$). Symmetry was broken by the relative motion of the oil with respect to the droplet crystal. Thus, we obtained a flowing 1D crystal of droplets that can move in 2D. The crystal exhibited visible longitudinal and transversal fluctuations, which were reminiscent of solid-state phonons (we henceforth term these normal modes 'phonons') (Fig. 1c,d and Supplementary Information, Movies S2–S4). We explored these modes by measuring their wave dispersion relations (Fig. 1e–h). This was done by tracking the positions of droplets in time and applying a Fourier transform to obtain the power spectrum of vibrations in terms wavevector $k$ and frequency $\omega$ (see the Methods section). We then extracted the dispersion relations of waves in the crystal, $\omega(k)$.

Surprisingly, the dispersion relations reveal the existence of acoustic phonons that propagate in the crystal at ultra-low frequencies of a few hertz. Manifestly, at Re $\sim 10^{-4}$, collective modes at such low frequencies cannot arise from inertial effects and are probably due to hydrodynamic interactions within the crystal. The main feature of the dispersion is an unusual sine-like curve that spans the Brillouin zone ($0 \leq k \leq \pi/a$) and has unique properties (Fig. 1). The linear behaviour of the curve $\omega(k) = C_{\rm s}k$ around $k = 0$ shows that these waves are acoustic and propagate at a sound velocity of $C_{\rm s} = (\partial\omega/\partial k)_{k\to 0} \approx 250$ µm s$^{-1}$. This velocity is some six orders of magnitude slower than sound in common liquids. Close to the edge of the Brillouin zone, $k = \pi/a$, the acoustic waves travel in the opposite direction at a velocity $-C_{\rm s}/2$, with a crossover between positive and negative group velocities. The longitudinal (Fig. 1e,g) and transversal (Fig. 1f,h) modes are identical in form and magnitude but travel in opposite directions $\omega_y(k) = -\omega_x(k)$. Dispersion of modes in the hydrodynamic crystal of droplets is markedly different from that of a harmonic crystal, where each wavevector has both symmetric forward and backward waves, $\omega(k) = \omega(-k)$, with standing waves at the edge of the Brillouin zone. In the hydrodynamic crystal, this symmetry is broken by the flow field of the oil such that waves travel only in one direction per wavevector. In addition, a standing wave appears at the group velocity crossover point within the Brillouin zone. A secondary feature of $\omega(k)$ is a straight line, $\omega = -u_{\rm d} \cdot k$, where $u_{\rm d}$ is the velocity of droplets relative to the channel. This simply stems from stationary defects along the channel that appear as if moving backwards at $-u_{\rm d}$, as the camera is moving in frame with the droplets. As we now explain theoretically, the unusual dispersion of the moving crystal arises from hydrodynamic interactions between droplets induced by the symmetry-breaking flow field.

The motion of each droplet perturbs the flow of the surrounding oil. This perturbation affects the other droplets in parabolic profile along the $z$ axis[14], and a potential flow in the $xy$ plane, such that the potential satisfies the Laplace equation. By demanding zero mass flux through the edge of the droplet[21], we find that the potential $\phi(\mathbf{r})$ the droplet induces is that of a two-dimensional dipole[23–25]:

$$\phi_d(\mathbf{r}) = R^2 \left(u_{oil}^\infty - u_d\right) \frac{\mathbf{r}\cdot\hat{x}}{r^2},$$

where $u_{oil}^\infty$ is the oil velocity far form the droplet. We visualized the oil streamlines $u_{oil}^\infty \hat{x} + \nabla\phi_d$ with fluorescent micro-beads and superimposed on them the dipole velocity field $\nabla\phi_d$ (Fig. 2a). Droplets interact with each other through the drag force[13], which is directed along the dipole velocity field (Fig. 2b). The drag that the $j$th droplet exerts on the $i$th droplet is given by $\xi_d \nabla\phi_d(\mathbf{r}_i - \mathbf{r}_j)$, a long-range force[22–24] that scales as $r^{-2}$, with a drag coefficient $\xi_d \simeq 8\pi\eta R^2/h$ ($\eta$ is oil viscosity). The potential of the crystal, as a dipole chain[26], is approximated by superposition of the single-droplet potentials $\phi(\mathbf{r}) = u_{oil}^\infty \mathbf{r}\cdot\hat{x} + \sum_j \phi_d(\mathbf{r} - \mathbf{r}_j)$, where the droplets are located in their crystal positions $\mathbf{r}_j = j\cdot a\cdot\hat{x}$. In these positions, droplets apply forces on each other that are all directed opposite to the oil velocity and do not cancel out (Fig. 2c). This is in contrast to harmonic crystals, where particles have equilibrium positions at which there is no net force.

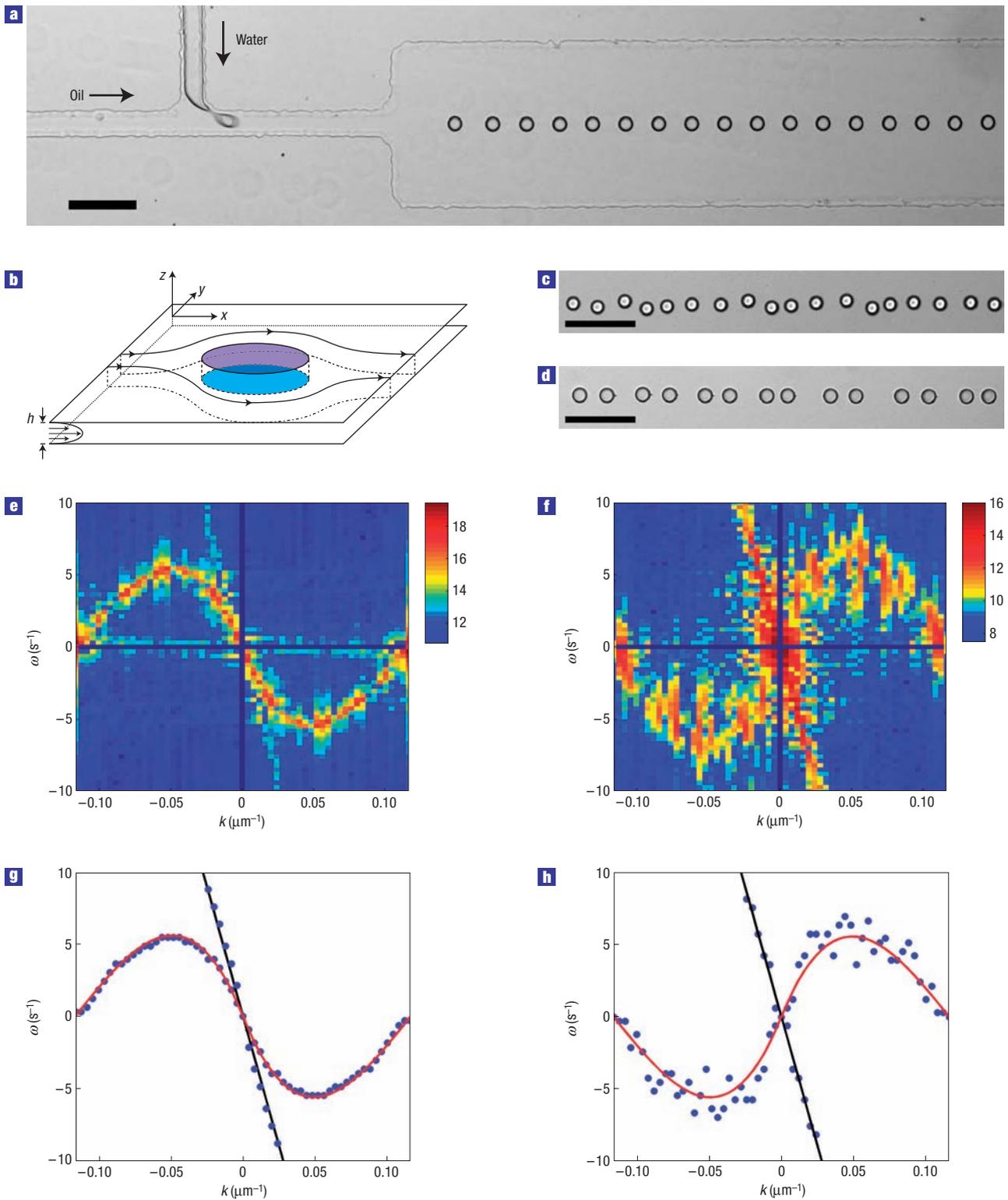

**Figure 1 Phonons in a 1D crystal of droplets. a**, A microfluidic device for generating a 1D flowing crystal of water-in-oil droplets. Uniform droplets form at the T-junction between water and oil (+surfactant) channels. The channel height was 10 μm. The width was 35 μm at the T-junction and 250 μm along the output channel. The scale bars are 100 μm. **b**, A droplet as a rigid disc between two plates, $h$ apart. Oil flows along the $x$ direction. The velocity profile is parabolic along the $z$ direction and a potential flow in the $x$–$y$ plane. **c,d**, Images of transversal and longitudinal acoustic waves. Fabrication defects seemed to increase the amplitude of oscillations. **e,f**, Intensity plot of the logarithm of the power spectrum of longitudinal (**e**) and transversal (**f**) waves as a function of wavevector and frequency $(k, \omega)$. Crystal spacing was $a = 27$ μm, droplets radius $R = 10$ μm, droplets velocity $u_d = 360$ μm s$^{-1}$ and oil velocity $u_{oil} = 1{,}730$ μm s$^{-1}$. **g,h**, The dispersion relations $\omega(k)$ are the peaks of the power spectrum (blue dots) showing two branches. (1) A skewed sine-like curve is due to the hydrodynamic interaction between droplets. The red line marks the theoretical result without adjustable parameters. (2) A straight line $\omega(k) = -u_d k$ (black) corresponds to relative motion of defects.

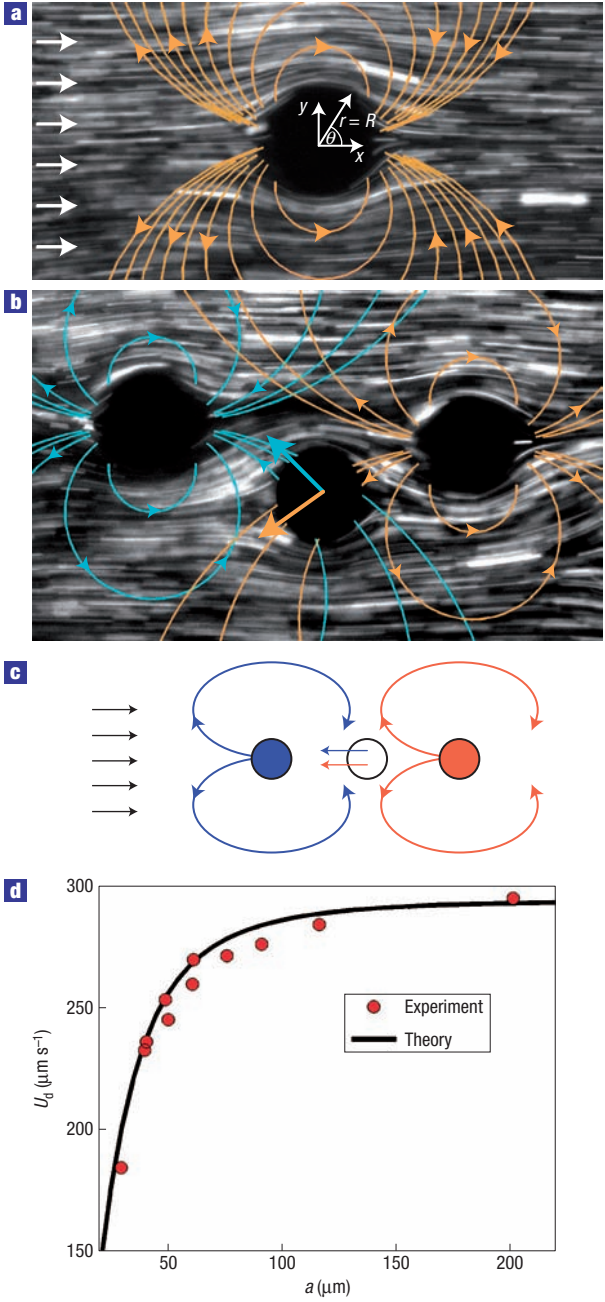

**Figure 2 Flow and hydrodynamic interactions between droplets. a**, Streamlines of the flow around a single droplet. The direction of the oil flow is indicated by white arrows. The orange lines show the dipole velocity field $\nabla\phi_d$. **b**, Interaction between droplets (bold arrows): the drag forces exerted on the middle droplet by its two neighbours. Force is directed along the corresponding dipole field, $\mathbf{F} \propto \nabla\phi_d$. **c**, In the absence of fluctuations, all interaction forces are directed opposite to the oil velocity. **d**, The 'peloton effect': a measurement of the velocity of droplets as a function of crystal spacing $a$ (red circles), for a given oil velocity $u_{oil}^\infty \approx 1{,}090\,\mu\text{m s}^{-1}$. The isolated droplet velocity was $u_d^\infty \approx 295\,\mu\text{m s}^{-1}$ and the friction coefficient $\mu \approx 24\,\text{mg s}^{-1}$. The solid line shows the theoretical results with no adjustable parameters.

Besides the drag force, the droplets are subject to a force of friction with the channel floor and ceiling. The friction induces a treadmill flow inside the droplet[4] and thus dissipates energy at a rate $\dot{\varepsilon} \sim \eta \int (\nabla\mathbf{v})^2 dV$, where $\mathbf{v}$ is the water velocity and $V$ is the droplet's volume[14]. If the flow pattern inside the droplet does not vary much and the only dependence on the droplet velocity is via scaling of the magnitude, $\mathbf{v}(\mathbf{r}) \sim u_d$, the dissipation rate is proportional to $u_d^2$, $\dot{\varepsilon} = \mu u_d^2$, where $\mu$ is the friction coefficient. By equating the work done by the friction force, $F_f u_d$, to the dissipation rate, we find that the friction is proportional to the droplet velocity, $F_f = \mu u_d$. This phenomenological law is verified by the measured dispersion relations and drag reduction reported below.

As inertia is negligible, drag and friction must balance out: $\mu u_d = \xi_d(u_{oil} - u_d)$, where $u_{oil}$ is the velocity of oil at the droplet's position. This gives $u_d = (1 + \mu/\xi_d)^{-1} u_{oil} = (u_d^\infty/u_{oil}^\infty)\, u_{oil}$, in which we calibrated the drag and friction coefficients on an isolated droplet flowing at velocity $u_d^\infty$. The equation of motion of the $n$th droplet is, therefore: $\dot{\mathbf{r}}_n = (u_d^\infty/u_{oil}^\infty) \sum_{j\ne n} \nabla\phi_d(\mathbf{r}_n - \mathbf{r}_j)$. Expanding in small deviations from crystal positions $(x_n, y_n) \ll a$, we find the wave equations of the crystal:

$$\begin{aligned}
\dot{x}_n &= (3C_s/\pi^2 a)\sum_{j=1}^{\infty}(x_{n+j} - x_{n-j})/j^3 \\
\dot{y}_n &= -(3C_s/\pi^2 a)\sum_{j=1}^{\infty}(y_{n+j} - y_{n-j})/j^3
\end{aligned} \quad (1)$$

with the sound velocity $C_s \simeq (2\pi^2/3)(u_d^\infty/u_{oil}^\infty)(u_{oil}^\infty - u_d^\infty)(R^2/a^2)$. The ultra-low sound velocity reflects the compressibility of the crystal and is set by the relative velocity of droplets with respect to the oil. For long wavelengths, equation (1) reduces to simple first-order wave equations $\partial x_n/\partial t \approx C_s \partial x_n/\partial x$ and $\partial y_n/\partial t \approx -C_s \partial y_n/\partial x$. Long-wavelength longitudinal modes move opposite to the flow, whereas transversal ones move with the flow, both at $C_s$. Substituting a plane wave in (1) gives the dispersion relations of phonons in the crystal:

$$\omega_x(k) = -\frac{6C_s}{\pi^2 a}\sum_{j=1}^{\infty}\frac{\sin(jka)}{j^3};\; \omega_y(k) = -\omega_x(k).$$

These dispersion relations are plotted along with the experimentally measured ones showing excellent agreement without any adjustable parameters (Fig. 1g,h). The peculiar properties of the dispersion relations noted above are a result of the symmetry-breaking the crystal and thus mediates hydrodynamic interaction[13]. As the droplet is a thin disc (Fig. 2a), the flow around it is that of the Hele–Shaw cell[21,22]: the flow is decomposed into a Poiseuille-like flow field. In particular, the skewness of the sine-like dispersion curve stems from the summation over interactions with distant neighbours. This proves that the hydrodynamic interactions are truly long range, extending over hundreds of micrometres. We note that the acoustic phonons reported here are absent from 1D systems with inertial forces and a trapping potential, where a non-acoustic 'sloshing' mode appears at $k \to 0$. Such systems include colloid particles trapped within laser beams[27,28] and dusty plasma particles trapped by an static electric field[29].

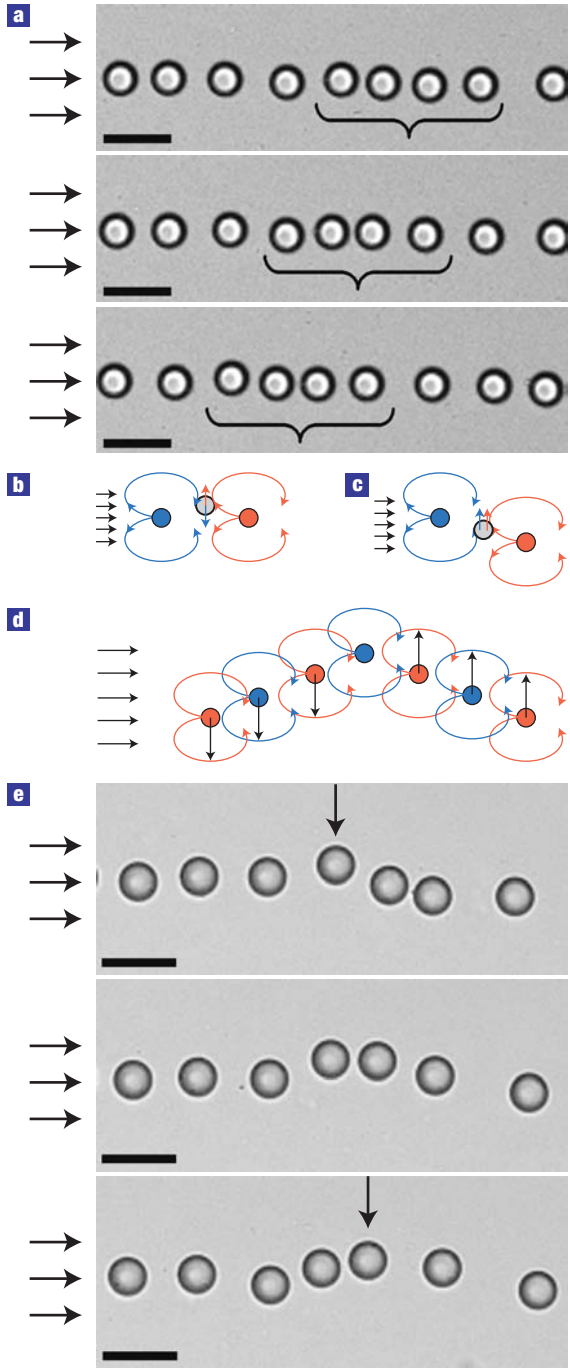

**Figure 3 Waves arise from dipole-like hydrodynamic interactions.**
**a**, Subsequent snapshots following a pack of nine droplets—a longitudinal travelling wave propagating opposite to the velocity of oil (black arrows). The scale bars are 50 μm. **b,c**, Nearest-neighbour interactions along the transversal direction—transversal forces on the middle droplet cancel out (**b**), or add up (**c**), depending on the triplet configuration. **d,e**, A triangle-like transversal wave propagating along the velocity of oil. The three snapshots (**e**) following a pack of seven droplets show the propagation of the peak (marked by an arrow). In (**d**), the right edge of the triangle moves up and the left edge moves down. As a result, the peak moves to the right—a travelling wave.

A peculiar consequence of the hydrodynamic interactions is that, even in the absence of vibrations, the crystal flow velocity depends on its spacing (Fig. 2d). This effect is quantified by summing the forces acting on a droplet due to the flow field of all the other droplets in the crystal, $\mathbf{u}_d = (\mu/\xi_d + 1)^{-1}(u_{oil}^\infty \hat{x} + \sum_j \nabla \phi_d(\mathbf{r} - \mathbf{r}_j))$. We find that $u_d(a)$ increases with $a$ as: $u_d(a) = u_d^\infty(1 + (\pi^2/3)(R/a)^2(u_{oil}^\infty - u_d^\infty)/u_{oil}^\infty)^{-1}$. Hence, with respect to the oil, the crystal moves faster as $a$ decreases, a result known as collective drag reduction. This is similar to sedimenting particles as well as to a pack of cyclists (peloton), where riding closely helps them to reduce drag and ride faster. This 'peloton effect' provides an intuitive explanation for longitudinal travelling waves. Consider a crystal in which a pack of droplets are more densely spaced than the remaining crystal (Fig. 3a and Supplementary Information, Movie S3). Owing to drag reduction, this pack moves slower than the rest. Droplets behind the pack catch up, whereas droplets ahead of it escape. The denser part of the crystal, therefore, travels opposite to the flow. Similarly, we can explain the origin of transversal waves by considering the dipole force fields of the droplets. When two droplets neighbouring a test droplet deflect symmetrically, the forces cancel out, whereas they add up for antisymmetric neighbouring droplet deflections (Fig. 3b,c). Therefore, combining such configurations into a triangular disturbance shows that the interaction between droplets pushes the two triangle edges in opposite directions, such that the triangle peak moves along the flow (Fig. 3d,e and Supplementary Information, Movie S4).

The 1D microfluidic crystal is susceptible to instabilities and we identified three types of them. In the prevalent type, the crystal breaks on a local fluctuation that grows in the middle of the crystal far from the droplet-formation zone (Fig. 4a and Supplementary Information, Movie S1). The structure of $1 + 3$ droplets repeats in many experiments and it is essentially a result of a non-linearity, which arises both from the large amplitude and from the interaction between longitudinal and transversal modes. A numerical solution of the full equations of motion, which are not restricted to small amplitudes and allow interaction between modes, reproduced the observed dynamics (see the Methods section). The second type of instability occurs at the droplet-formation zone on abruptly cutting the crystal by stopping water flow while maintaining oil flow. The drifting crystal then exhibits a wandering motion in the transversal direction and a longitudinal pairing wave that advances from the crystal trailing end (Fig. 4b,c and Supplementary Information, Movie S5). The third instability, which we term the 'zigzag mode', occurs close to the droplet-formation zone as the crystal bifurcates into two parallel crystals (Fig. 4d and Supplementary Information, Movie S6). The propensity to zigzag is enhanced near the droplet-formation zone where the asymmetry of the crystal causes transient force imbalance and thus growth of the zigzag amplitude (Fig. 4e,f). Finally, we note that waves and instabilities at ultra-low Re have been observed in the context of elastic turbulence[30]. It requires, however, the nonlinear response of a non-newtonian polymer solution, whereas our acoustic phonons occur in simple newtonian oil. The present two-phase microfluidic system offers an additional example of pattern formation, in which nonlinear long-range interactions combine with the inherently linear low Reynolds number flow.

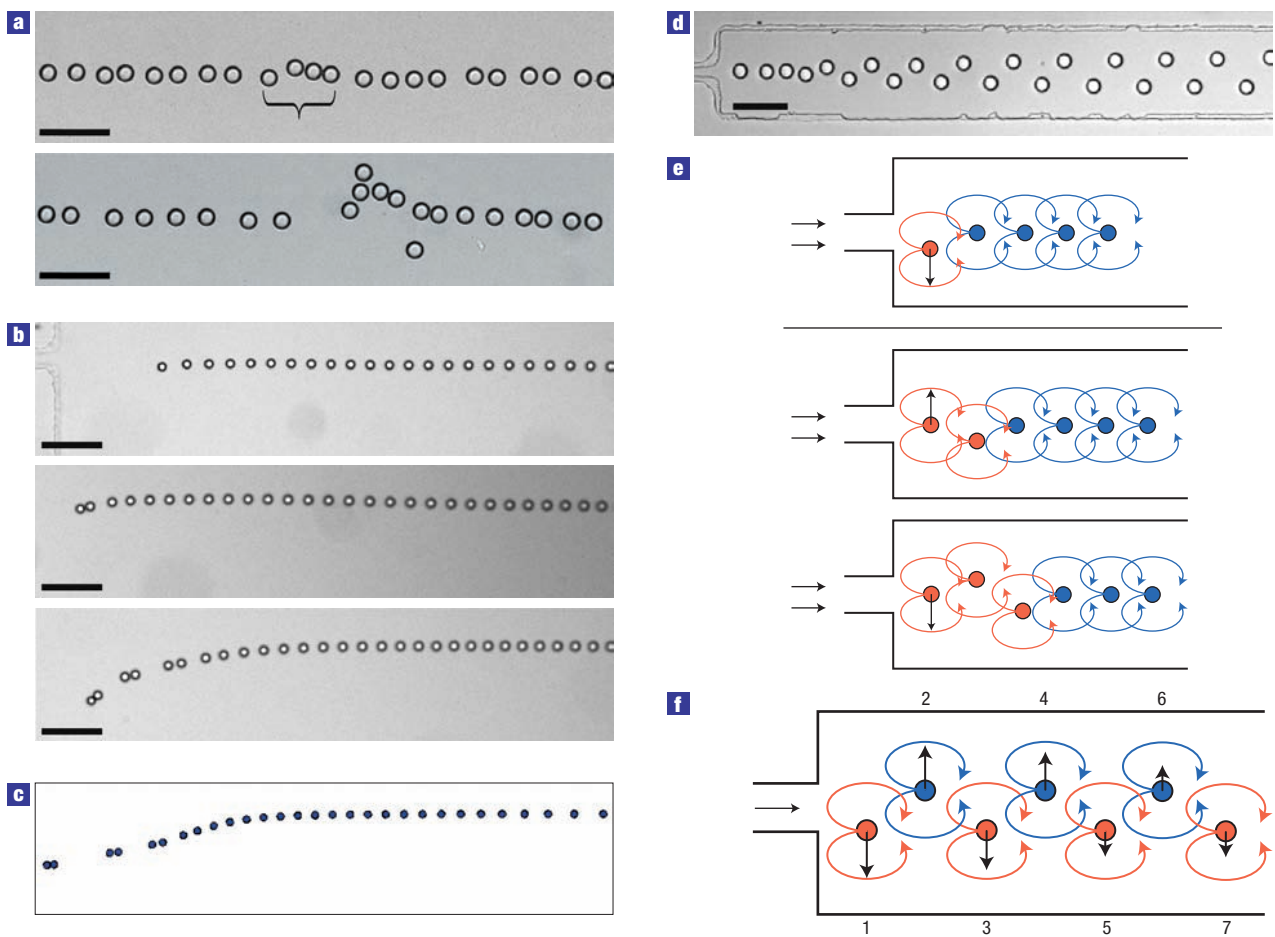

**Figure 4** Crystal instability. **a**, Subsequent images of a (1 + 3; marked) local fluctuation that grows and leads to crystal instability. **b**, Instability at the trailing end of the crystal on stopping droplet formation, showing transversal wandering and a pairing wave. Transversal wandering: a deflection at the last droplet grows because it is pushed in the same direction by all other droplets, with no neighbours behind it to cancel these forces. Concomitantly, the last droplet pulls the droplet ahead of it off the crystal axis. This deflection advances and the entire crystal drifts aside. The pair formation results from the 'peloton effect': the last droplet has fewer neighbours to slow it down and, therefore, catches up with the droplet ahead to form the first pair. The interaction within the pair slows it down with respect to the crystal. When the pair lags further behind, the third droplet becomes effectively the last droplet, thus a cascade of pair-formation ensues. **c**, This instability was also obtained in simulation. **d**, Zigzag instability close to the droplet formation area. **e**, The asymmetry at the trailing end of the crystal makes it a source of zigzag waves. The scheme shows a small deflection of the trailing (orange) droplet that increases due to transient force imbalance near the formation area. When the next droplet emanates, it is pushed in the opposite direction, and so on (arrows mark transversal forces). **f**, Given a zigzag wave near the formation area, trailing-end asymmetry causes the zigzag to grow: the droplet at the trailing end (1) is pushed downwards by droplets 2, 4, 6, . . ., because there are no droplets on the left to cancel out the transversal force they exert on droplet 1. Following the same logic, droplet 2, in turn, is pushed upwards by droplets 5, 7, 9, . . ., and so on.

## METHODS

### DEVICE FABRICATION

The microfluidic device had two inlets and one outlet, all connected at a T-junction (Fig. 1a and Supplementary Information, Movie S1). Droplets emanated at a constant rate with uniform radii and inter-droplet distances, due to a balance between the tearing shear force and the stabilizing surface tension[2]. Channels height was 10 µm. Water and oil inlets were 100 µm in width and 3 mm long. T-junction width was 35 µm and output channel widths ranged from 175 µm to 500 µm. Output length was either 1 cm or 2 cm. The microfluidic device was fabricated using a standard soft-lithography technique[31]. Reusable moulds were made from SU-8-2010 negative photoresist (Microchem). Channels were prepared by casting of PDMS elastomer (poly di-methyl siloxane, Sylgard-184, Dow-Corning) on the mould and curing at 80 °C for 40 min. After separating the cured PDMS from the mould and punching holes at the inlets and outlet, the PDMS was irreversibly attached to a clean glass that constituted the channel floor. Irreversible sealing was done by oxidizing both surfaces in a plasma asher (March Plasmod) for 50 s at 150 W before attachment. We also used devices with a PDMS floor, rather than glass, which gave similar results.

### MATERIALS

We used light mineral oil (Sigma, M5904, viscosity 30 cp) with 2% (w/w) span-80 surfactant (Sigma) and deionized double-distilled water (Millipore, 18 MΩ cm). Fluids were kept in 1 ml plastic syringes and connected to the device by tygon tubes. Fluids were driven by nitrogen pressure typically between 4 and 20 psig, where water pressure was usually 60–80% of the oil pressure. Flow was visualized (Fig. 2) by adding fluorescent polystyrene micro-beads to the oil (diameter 0.4 µm).

### EXPERIMENT

The device was mounted on an inverted microscope (Olympus, IX-71) with a motorized stage (Newport UMR5.25 and CMA-25CC). After oil and water flow was stabilized to give a uniformly spaced crystal of droplets, the flow was recorded using a digital camera (PCO, SensiCam-QE). In the 'peloton effect' measurements the device was static. The crystal constant $a$ was set by small changes in water pressure keeping the oil velocity essentially constant. In the power-spectrum measurements of the vibrations in the crystal, the sampling time had to be increased to get a sufficient resolution in $\omega$. This was done by tracking a pack of droplets in motion by moving the device with respect to the microscope using the motorized stage. In the experiment shown in Fig. 1, we followed the motion of a pack of ∼60 droplets for ∼20 s over a distance of ∼1 cm.

### IMAGE PROCESSING

The sequence of images acquired in the experiment was analysed using a precise tracking algorithm (the Moses–Abadi algorithm, the 'Abadiscope'; unpublished), implemented in Matlab. The algorithm locates the droplets in an image, finds the positions of their centres and tracks the trajectory of each droplet in time. To extract the spectrum of crystal vibrations, it is crucial to accurately determine the positions of the droplets' centres. The algorithm works, therefore, in two stages. First, the location of the droplet is determined to low accuracy on the basis of the contrast between the droplet boundary and the background (binary threshold). Next, the centre is found with higher accuracy by calculating an optimal fit between the image of the droplet and a ring with a gaussian intensity profile along its radius. This fit has three parameters: the two coordinates $(x, y)$ of the centre and the droplet radius. The width of the gaussian cross-section was determined manually. Using this method, it is possible to locate the coordinates of a droplet centre position to sub-pixel accuracy.

### DATA ANALYSIS

The power spectrum of crystal vibrations was calculated from the coordinates $[x(n,t), y(n,t)]$ of all the droplets $n=1\ldots N_d$ at all times $t=1\ldots N_f$. To apply a Fourier transform, the deviations of droplets from their crystal positions need to be found. In the $y$ direction it is straightforward, because the crystal is, by definition, at $y=0$. In the $x$ direction, however, it is hard to determine the positions of the moving vibrating crystals. For this purpose we defined a new $x$ coordinate, which is the difference between adjacent droplets:

$$\xi(n,t) = \begin{cases} 0, & n=1 \\ x(n,t) - x(n-1,t) & \text{otherwise.} \end{cases}$$

We used a discrete Fourier transform (Matlab) both in space and in time:

$$\tilde{X}(k,\omega) = \sum_{n=1}^{N_d}\sum_{t=1}^{N_f} \xi(n,t) e^{-(2\pi i/L)(k-1)(n-1)a} e^{-(2\pi i/T)(w-1)(t-1)\Delta t}$$
$$\tilde{Y}(k,\omega) = \sum_{n=1}^{N_d}\sum_{t=1}^{N_f} y(n,t) e^{-(2\pi i/L)(k-1)(n-1)a} e^{-(2\pi i/T)(w-1)(t-1)\Delta t},$$

where $\Delta t$ is the time between adjacent frames, $a$ is the equilibrium distance between droplets, $T = N_f \Delta t$ is the overall measurement time and $L = N_d a$ is the total length of the sublattice that we tracked. Note that here $k, \omega, n$ and $t$ are all indices. The wavenumbers and frequencies are $2\pi k/L$ and $2\pi\omega/T$, respectively. The longitudinal and transversal dispersion relations of waves in the crystal, $\omega(k)$, were obtained as the peaks of the power spectrum $|\tilde{X}(k,\omega)|^2$ and $|\tilde{Y}(k,\omega)|^2$ of the fluctuations, such that for each $k$ we found the $\omega$ that corresponds to the power-spectrum peak. In this process, we disregarded the high-intensity lines around the $k=0$ and $\omega=0$ lines in the power spectrum.

### SIMULATION

Our simulation is based on a numerical solution of the general equations of motion for the droplets lattice: $\dot{\mathbf{r}}_n = (u_d^\infty/u_{oil}^\infty)\nabla\phi(\mathbf{r}_n)$. This is in contrast to the linearized form (equation (1)) that is valid only for small fluctuations. For a lattice of $N$ droplets, we obtain a set of $2N$ coupled first-order ordinary differential equations:

$$\dot{x}_i = \Lambda \sum_{j\neq i} \frac{y_{ij}^2 - x_{ij}^2}{(x_{ij}^2+y_{ij}^2)^2}$$
$$\dot{y}_i = \Lambda \sum_{j\neq i} \frac{-2y_{ij}x_{ij}}{(x_{ij}^2+y_{ij}^2)^2}$$

where $\Lambda \equiv R^2(u_{oil} - u_d)u_d/u_{oil}$, $x_{ij} \equiv x_i - x_j$ and similarly $y_{ij} \equiv y_i - y_j$. We work in a frame of reference in which the crystal is at rest, thus the droplet formation area moves at velocity $-u_d$. To fully model the experiment, new droplets are added at the formation area at time intervals $a/u_d$. We start with an initial lattice of $N = 100$–$200$ droplets set at crystal positions $(x_i, y_i) = (a \cdot i, 0)$ optionally with a superimposed initial perturbation. The equations are solved using Matlab ODE solver.

### Acknowledgements
We thank G. Falkovich, S. Fleishman, A. Libchaber, E. Moses, S. Safran, V. Steinberg and D. Weitz for useful suggestions and discussions. R.B.-Z. is an incumbent of the Beracha career development chair. This work was supported by a grant from the Israel Science Foundation grant (Bikura) to R.B.-Z. Correspondence and requests for materials should be addressed to R.B.-Z.